\documentclass[english,aps,twocolumn,prr,showpacs,superscriptaddress]{revtex4-1}
\usepackage[T1]{fontenc}
\usepackage[latin9]{inputenc}
\usepackage{babel}
\usepackage{graphicx}
\usepackage{bm}
\usepackage{amsmath}
\usepackage{amssymb}
\usepackage{amsfonts}
\usepackage{dcolumn}
\usepackage{bbold}
\usepackage{tikz}
\usepackage{siunitx}
\usepackage[margin=15mm]{geometry}
\colorlet{linkequation}{blue}
\usepackage[colorlinks=true,linkcolor=blue,citecolor=blue,urlcolor=blue]{hyperref}
\usetikzlibrary{shadings}

\RequirePackage[normalem]{ulem} %DIF PREAMBLE
\RequirePackage{color}\definecolor{RED}{rgb}{1,0,0}\definecolor{BLUE}{rgb}{0,0,1} %DIF PREAMBLE
\usepackage{color}

\renewcommand\vec[1]{\bm{{#1}}}
\newcommand{\uvec}[1]{\vec{\hat{#1}}}

\begin{document}

\title{Terahertz spin dynamics driven by an optical spin-orbit torque}

\author{Ritwik Mondal}
\email[]{ritwik.mondal@uni-konstanz.de}
\affiliation{Fachbereich Physik, Universit\"at Konstanz, DE-78457 Konstanz, Germany}
\author{Andreas Donges}
\affiliation{Fachbereich Physik, Universit\"at Konstanz, DE-78457 Konstanz, Germany}
\author{Ulrich Nowak}
\affiliation{Fachbereich Physik, Universit\"at Konstanz, DE-78457 Konstanz, Germany}

%------------------------------------------------------------------------
\begin{abstract}
Spin torques are at the heart of spin manipulations in spintronic devices.
Here, we examine the existence of an optical spin-orbit torque, a relativistic spin torque originating from the spin-orbit coupling of an oscillating applied field with the spins. 
We compare the effect of the nonrelativistic Zeeman torque with the relativistic optical spin-orbit torque for ferromagnetic systems excited by a circularly polarised laser pulse.
The latter torque depends on the helicity of the light and scales with the intensity, while being inversely proportional to the frequency. Our results show that the optical spin-orbit torque can provide a torque on the spins, which is quantitatively equivalent to the Zeeman torque. 
Moreover, temperature dependent calculations show that the effect of optical spin-orbit torque decreases with increasing temperature.
However, the effect does not vanish in a ferromagnetic system, even above its Curie temperature.

\end{abstract}

%-------------------------------------------------------------------------
\maketitle
\date{\today}

%{\red TODO: discussion of the limit $\omega\to0$, analogy to IFE Ref. ?}

%{\it Introduction:}
\section{Introduction}
Interest in controlling spins by means of circularly polarised pulses has grown immensely due to its potential applications in spin-based memory technologies \cite{Kimel2007-Review, Stanciu2007, Kimel2005}. 
Apart from the heat-assisted spin manipulations, the spins can {\it also} be controlled using  the inverse Faraday effect (IFE) \cite{John2017,Mangin2014,lambert14}, the magnetic field of the terahertz pulses \cite{Donges2018,Kampfrath2010,Kampfrath2013,Wienholdt2012PRL} and an optical spin-orbit torque (OSOT) that does not impart angular momentum into the spin system \cite{tesarova13,Choi2020}.    
To be able to explain such effects theoretically, one has to simulate spin dynamics including several nonrelativistic and relativistic effects that might appear at ultrashort timescales  
\cite{tesarova13,Kampfrath2013,Kampfrath2010,Wienholdt2012PRL,Choi2020}.

The Landau-Lifshitz-Gilbert (LLG) equation of motion, consisting of precession of a spin moment around an effective field and a transverse relaxation, has been used extensively in the past to simulate such spin dynamics \cite{landau35,gilbert04,kronmuller2003micromagnetism}.  
However, for a spin system excited by ultrashort laser pulses, a stochastic LLG equation of motion with atomistic resolution is required due to the strong thermal fluctuations in order to study the dynamical processes \cite{Kazantseva2007,nowak2007handbook,Evans2014,Wienholdt2013,Frietsch2020}. In these equations, a stochastic field is added to the effective field, in order to quantify the thermal fluctuations in the spin system. 
Nonetheless, at ultrashort timescales, the exact form of the LLG equation of motion has to be questioned as several other relativistic phenomena can occur. 
Therefore, we seek for an equation of motion that can capture all the possible interactions during ultrashort laser excitations of a spin system.

In a previous work, starting from the relativistic Dirac Hamiltonian yet including the magnetic exchange interactions, a rigorous derivation of the LLG equation of motion has been provided \cite{Mondal2016}. 
To treat the action of a laser pulse and the corresponding interactions, the Dirac-Kohn-Sham equation with external magnetic vector potential was considered \cite{Mondal2015a,Mondal2015b,crepieux01}. 
To this end, a semirelativistic expansion of the Dirac-Kohn-Sham Hamiltonian that includes several nonrelativistic and relativistic spin-laser coupling terms was derived \cite{Mondal2017JPCM}. 
Having these coupling terms, an extended equation of motion that includes {\it not only} the spin precession and damping, {\it but also} other relativistic torque terms was obtained \cite{Mondal2018PRB}. 
One of these torque terms is the field-derivative torque which appears due to the time-dependent field excitation, e.g., in case of THz pulse excitation \cite{Mondal2019PRB}. 
Another new torque term is the OSOT, which stems from the spin-orbit interaction of the applied field with the electron spins i.e., it imparts spin-angular momentum of the applied field to the spins.    
The new equation of motion for the reduced
 magnetization vector $\bm{m}_i(t)$ including OSOT cast in the form of an LLG equation is  
\begin{align}
    \frac{\partial\bm{m}_i}{\partial t}  = & 
     - \gamma \bm{m}_i\times \left(\bm{B}^{\rm eff}_i + \bm{B}_{\rm OSOT}\right)\nonumber\\
     & + \bm{m}_i \times\left(\mathcal{D}\cdot \frac{\partial\bm{m}_i}{\partial t} \right) \,.
     \label{LLG_equation}
\end{align}
We define the gyromagnetic ratio as $\gamma$ and $\mathcal{D}$, representing the damping parameter, is in general a tensor. For simplification, we consider only a scalar damping parameter $\alpha$ that can be expressed as $\alpha = \frac{1}{3}{\rm Tr}(\mathcal{D})$ \cite{Mondal2016}. 
The effective field is represented as $\bm{B}^{\rm eff}_i$, which is the derivative of the total magnetic energy without the relativistic light-spin interaction with respect to $\bm{m}_i$ as will be specified later on in detail.
The additional field $\bm{B}_{\rm OSOT}$ describes the opto-magnetic field induced by the laser pulse due to the OSOT phenomenon.    

In this article, we investigate the effect of the OSOT term within atomistic spin dynamics simulations. 
First, we revisit the derivation of the OSOT from a general spin-orbit coupling Hamiltonian that has been derived within a relativistic formalism \cite{Mondal2016}. 
We find that the OSOT depends on the helicity, frequency and intensity of the light pulse. 
If the intensity is high and frequency is low, we expect the OSOT terms to show the most significant effects. 
We simulate the spin dynamics for a spin model, representative for bcc Fe, with the OSOT and find that the OSOT can provide significant contributions at the THz regime. 
Studying the temperature dependence, we find that the OSOT effects is robust against thermal fluctuations, i.e., we observe no significant reduction of the strength of the OSOT even up to the critical temperature.

\section{Optical spin-orbit torque}
\begin{figure}[hbt!]
\centering
\includegraphics[scale = 0.28]{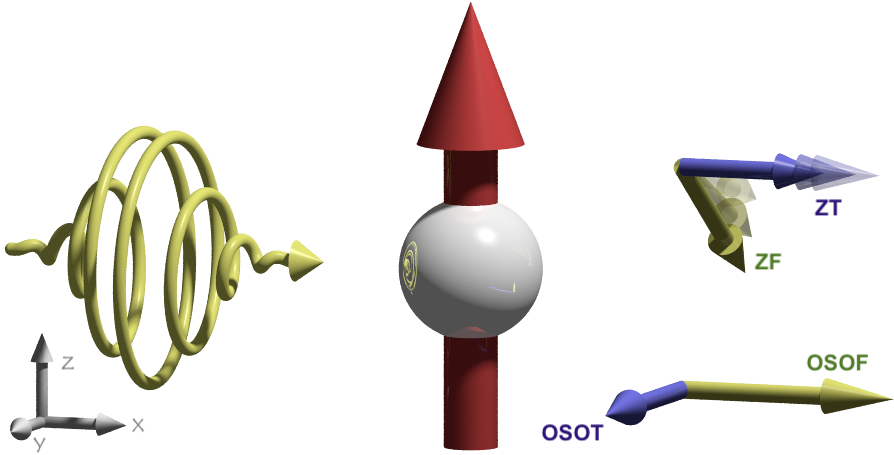}
\caption{(Color Online) ZT and OSOT for a single spin (big red arrow) excited by an elliptically polarised laser pulse. 
The fields and torques are represented by yellow and blue arrows, respectively. Due to the presence of elliptical polarisation, the ZF and ZT are drawn as blurry.} 
\label{Schematics}
\end{figure}
The generalized spin-orbit coupling (SOC) Hamiltonian, as derived within the fully relativistic Dirac framework, can be written as \cite{Mondal2015b,Mondal2016,Mondal2018PRB}
\begin{align}
    \mathcal{H}_{\rm SOC} & = -\frac{e\hbar}{8m^{2}c^{2}}\, \bm{\sigma}\cdot\left[ \bm{E}_{\rm tot}\times\left(\bm{p}-e\bm{A}\right)-\left(\bm{p}-e\bm{A}\right)\times\bm{E}_{\rm tot}\right],
    \label{SOC_eq1}
\end{align}
where the electric fields are represented as $\bm{E}_{\rm tot} = \bm{E}_{\rm int} + \bm{E}_{\rm ext}$ and  $\bm{E}_{\rm ext} = -\frac{\partial \bm{A} }{\partial t} - \bm{\nabla}\Phi$ with ($\bm{A}, \Phi$) as magnetic vector and scalar potentials. 
%The constant prefactor is $\kappa = \frac{e\hbar}{8m^{2}c^{2}}$, where 
The physical constants have their usual meanings and $\bm{\sigma}$ denotes the electron's spin through Pauli spin matrices and $\bm{p}$ is the electron momentum.  

Note that the SOC can occur in several ways, as described in the following: (i) the angular momentum of an electron couples to the spin of the electron --- this can be expressed as $\bm{\sigma}\cdot (\bm{E}_{\rm int} \times \bm{p})$, (ii) the first-order magnetic vector potential of the electromagnetic (EM) field couples to the spins --- this can be expressed as $\bm{\sigma}\cdot (\bm{E}_{\rm int} \times \bm{A})$, (iii) the spin angular momentum of the EM field couples to the spin --- this can be expressed as $\bm{\sigma}\cdot (\bm{E}_{\rm ext} \times \bm{A})$. 

%The first type of spin-orbit coupling is well-known in the literature. 
In a spherically symmetric potential, the first type SOC can be written as traditional $\bm{l}\cdot \bm{s}$ coupling, where $\bm{l}$ and $\bm{s}$ represent the orbital and spin angular momentum respectively. 
It provides explanations to several relativistic effects in magnetism e.g., magnetic Gilbert damping \cite{Mondal2016,Gilmore2007thesis,gilmore07} and many others \cite{Moriya1960,Dzyaloshinskii}.
%, Dzyaloshinskii-Moriya interaction and so on. 
Such a SOC exists even without an external field. In contrary, the latter two types of SOC depend explicitly on the external field and can be written as  

\begin{align}
\mathcal{H}_{\rm SOC}^\prime %& =  \frac{e^2\hbar}{8m^{2}c^{2}}\, \bm{\sigma}\cdot\left[ \bm{E}_{\rm tot}\times\bm{A}-\bm{A}\times\bm{E}_{\rm tot}\right]\nonumber\\
    & = \frac{e^2\hbar}{4m^{2}c^{2}}\, \bm{\sigma}\cdot\left[ (\bm{E}_{\rm int} +\bm{E}_{\rm ext} )\times\bm{A}\right]\,.
\end{align}
The total internal field depends on the intrinsic field $\bm{E}_{\rm int}^0$ that even exist without the applied field and the applied field itself. 
Within the linear response theory we write $\bm{E}_{\rm int}  = \bm{E}_{\rm int}^0 + \zeta \bm{E}_{\rm ext}$, where $\zeta$ defines the coupling strength of the applied EM field relative to the intrinsic one. We thus consider the optical spin-orbit coupling as 
\begin{align}
\mathcal{H}_{\rm OSOC} & = \frac{e^2\hbar(1+\zeta)}{4m^{2}c^{2}}\bm{\sigma}\cdot (\bm{E}_{\rm ext}\times\bm{A})\,.
\label{osoc}
\end{align}
Using the definition of Zeeman coupling of $-\mu_s \bm{\sigma} \cdot \bm{B}_{\rm OSOT}$, the optical spin-orbit coupling

can be shown to induce a field (see Appendix \ref{appendixA} for details)
\begin{align}
    \bm{B}_{\rm OSOT} & = \frac{e^2\hbar(1+\zeta)}{4m^{2}\mu_s}\frac{B_0^2}{\omega} \,\sin \eta\, \hat{\bm{e}}_x \,,
    \label{induced_OSOF}
\end{align}
where $B_0 (= E_0/c)$, $\eta$ and $\omega$ are the  envelope of the oscillating Zeeman field, helicity and angular frequency of the elliptically polarised light, respectively. 
$\mu_s$ defines the magnetic moment. %and $\zeta$ is the coupling between the EM field and spins. 
 Note, that due to the OSOT, the induced field points along the direction of the energy flux of the propagating wave. 
 Additionally, note that the field $\bm{B}_{\rm OSOT}$ is largest for circularly polarised light ($\eta = \pm \pi/2$). 
 However, if the coupling strength $\zeta$ is not the same for right and left circularly polarised light, their combined effect could lead to nonzero $\bm{B}_{\rm OSOT}$ for linearly polarised light.
{The parameter $\zeta$ depends on the electron density and absorption of the light that can be different for right and left circularly polarised light, leading to a magnetic circular  dichroism (MCD).}
In the following, we simulate the Zeeman effect from the electromagnetic THz field and the optical spin-orbit coupling effects simultaneously, and make a comparison between these two effects in a ferromagnetic system.

%{\it Atomistic spin model:}
\section{Atomistic spin simulations}
In order to compute the spin dynamics, it is convinient to transform the implicit form of our equation of motion to the explicit Landau-Lifshitz (LL) form. 
For a scalar damping parameters, $\alpha$, the LLG equation~\eqref{LLG_equation} can be recast as
\begin{align}
&\frac{\partial\bm{m}_i(t)}{\partial t} 
 = -\frac{\gamma}{(1+\alpha^2)}\bm{m}_i\times \left(\bm{B}^{\rm eff}_i + \bm{B}_{\rm OSOT}\right)\nonumber\\
& -  \frac{\gamma\alpha}{(1+\alpha^2)}\bm{m}_i\times \left[\bm{m}_i\times\left(\bm{B}^{\rm eff}_i  + \bm{B}_{\rm OSOT}\right)\right]\,.
\label{LL-equation}
\end{align}
The effective field, $\bm{B}^{\rm eff}_i$, is the derivative of total energy with respect to the magnetic moment, $\bm{B}^{\rm eff}_i = - \frac{1}{\mu_s^i} \partial\mathcal{H}/\partial \bm{m}_i$.
The LLG equation, Eq.~\eqref{LL-equation}, hereby consists of the so called field-like (see also Fig.~\ref{Schematics}) and the weaker damping-like torque which is  proportional to the Gilbert damping coefficient $\alpha=0.01$ and describes the coupling of the spins to a heat bath.

In the following we consider a spin model for bcc Fe. 
The total Hamiltonian of the system (without the relativistic spin-light coupling term) $\mathcal{H}$ can be expressed as
\begin{align}
    \mathcal{H}  = & - \sum_{i< j} J_{ij} \bm{m}_i\cdot \bm{m}_j - \sum_i d_z \left(m_{i}^z\right)^2 \nonumber\\
    & -\mu_s \bm{B}(t) \cdot\sum_i\bm{m}_i\,. 
\end{align}
The first term describes the traditional Heisenberg exchange energy, considering the exchange constants $J_{ij}$ up to the third nearest neighbour. 
The second term represents the uniaxial anisotropy with energy constant $d_z$ and the last term is the Zeeman coupling with a time-dependent field $\bm{B}(t) =\mathcal{R}\left[\bm{B}_0\,\exp\left({-\frac{t^2}{2\tau^2}} - i\omega t\right)\right]$. 
The ellipticity of the applied pulse is taken into account through $\bm{B}_0 = {\frac{B_0}{\sqrt{2}} \left(\uvec{y} + \mathrm e^{i\eta}\uvec{z}\right)}$, which considers the major and minor axes of the ellipse and $\tau$ defines the pulse duration.
In the following we use the abbreviations Zeeman field (ZF), Zeeman torque (ZT), and optical spin-orbit field (OSOF).  

\begin{figure*}[t!]
\centering
\includegraphics[width =1 \textwidth]{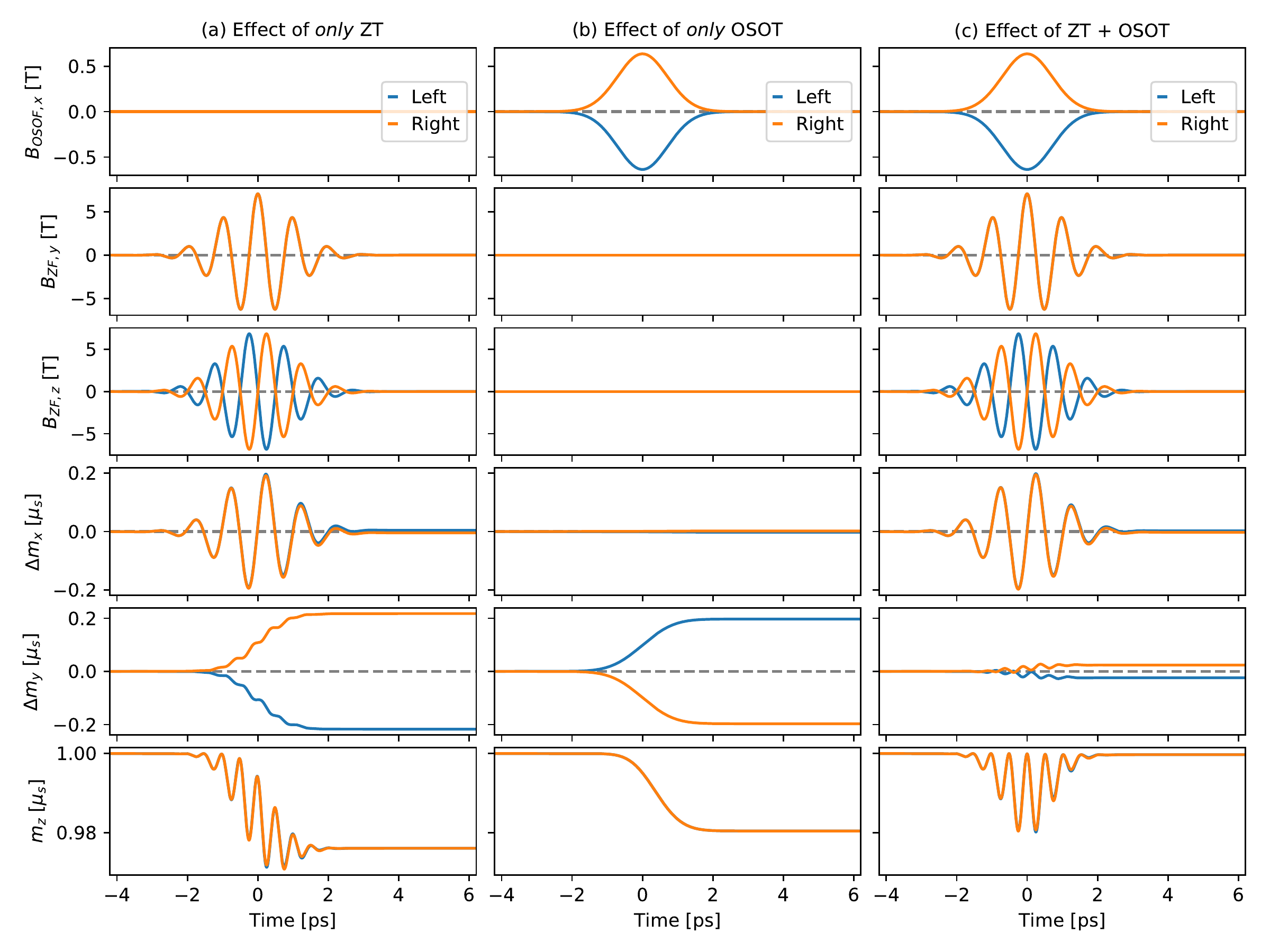}
\caption{(Color Online) The dynamical effects of ZT and OSOT for Fe at an { maximum} applied field of 10 T. The calculations include (a) {\it only} { ZT}, (b) {\it only} OSOT  and (c) both the ZT and OSOT. The first row shows the applied ZF which has $y$ and $z$ components, the second row represents the induced OSOF which has only an $x$-component. 
The other three rows show the change in magnetization along $x$, $y$ and $z$ components respectively.} 
\label{Ferro_fig1}
\end{figure*}

A striking difference between ZT and OSOT is that the ZT does not depend on the frequency of the pulse and is proportional to the amplitude of the applied field pulse. 
However, the OSOT depends inversely on the frequency and is proportional to the square of the field amplitude, i.e. the intensity. 
To quantify the OSOT effects, one can estimate its amplitude by computing the characteristic field constant of the OSOF
\begin{align}
\tilde B(\omega,\zeta) = \frac{4 m_e^2 \mu_s \omega}{e^2\hbar(1+\zeta)}.
\end{align}
Considering a central frequency of $f=\SI{1}{\tera\hertz}$, we find that $\tilde B(2\pi f, 0)=\SI{157}{\tesla}$.
Hence an electromagnetic field amplitude of about \SI{10}{\tesla} could induce a OSOF of around 1\,T, even for the low limit of the coupling strength, $\zeta \rightarrow 0$.
Needless to mention that the OSOT effects might exceed the ZT if $\zeta$ is higher. However, the results shown here were computed for the limit $\zeta \rightarrow 0$.
Such ultra-intense THz fields are currently only achievable with linear polarisation, with recent experiments pointing towards achieving circular polarized laser pulses in the terahertz frequency regime \cite{Jia2019,Jasper2020}.

%{\it Application to ferromagnets:} 
\section{Application to ferromagnets}
In our following simulations, we consider bcc Fe as a ferromagnetic sample.
For the zero-temperature simulations it is sufficient to solve the LLG equation for a single spin, due to the homogeneity of the THz pulse excitation.
The choice of exchange parameters is only relevant at elevated temperatures.
We shine an electromagnetic pulse having \SI{1}{\pico\second} pulse duration along the $\uvec{x}$-direction with $y$ and $z$ field components as shown in the Fig.~\ref{Schematics}. 
We will mainly discuss the limit $d_z\approx 0$, since the uniaxial anisotropy does not have a significant impact on the ultrafast spin dynamics (see Appendix~\ref{appendixB} for details).

\subsection{Zero temperature simulations}

In particular we compare the effect of ZT and OSOT at zero temperature. 
As the optical pulse is applied along the $\uvec{x}$-direction ($\bm{ k} = \vert \bm{k}\vert \hat{\bm{ x}}$), we calculate the change in magnetization for $y$ and $z$ components in Fig.~\ref{Ferro_fig1}(a). For the case of only ZT the induced OSOT remains obviously zero. 
For a maximum of \SI{10}{\tesla} applied ZF, the change in magnetization is about 50\,\% for the $y$-component and 5\,\% for the $z$-component. 
The reason for the triggered magnetisation dynamics is the following: the ZF from the optical pulse has two components $B_y$ and $B_z$. 
The $B_y$ component exerts a torque along $-\uvec{x}$ on the equilibrium spin direction: 
\begin{align}
\Delta \vec m \propto m_0\uvec{z} \times B_y\uvec{y} = -m_0B_y\uvec{x},\label{eq:dmx}
\end{align}
{with equilibrium magnetisation $m_0$.}
On the other hand, the $B_z$ component of the EM field does not exert any torque on the Fe spins at first as the spins are initially aligned along $\uvec{z}$. 
However, as soon as the above mentioned torque induced some magnetisation $\Delta m_x$, the $B_z$ component of the EM pulse, one quarter period later, exerts a torque along $\uvec{y}$ direction: 
\begin{align}
\Delta \vec m \propto \Delta m_x\uvec{x} \times B_z\uvec{z} = m_0B_yB_z\uvec{y}.\label{eq:dmy}
\end{align}
Therefore, there exists a superposition between two torques for the ZF.
Note that the change in  $\Delta m_y$ magnetization is antisymmetric in the helicity of light, which suggests that the effect is similar to IFE \cite{Pershan1966}. 
Further note that the \emph{ab initio} calculations of the IFE were calculated using a nonrelativistic theory \cite{Battiato2014,Berritta2016}. However, present theory is based on a relativistic interaction Hamiltonian. { According to Eq.\ (\ref{induced_OSOF}), the induced field diverges in the limit $\omega\rightarrow 0$, which can be explained by the previous theories of IFE \cite{Alireza2013,Popova2011,HERTEL2006L1}, and  is in accordance with the  {\it ab initio} calculations of IFE \cite{Battiato2014,Berritta2016,Frank2016}. }
We would also like to mention that these \emph{ab initio} calculations showed the IFE being asymmetric in helicity for ferromagnets which includes the absorption of light \cite{Berritta2016}.

The OSOF, on the other hand, has only one component $B_x$ along $\uvec{x}$ direction, see Fig.~\ref{Ferro_fig1}\,(b). 
Thus, it exerts a torque along $\uvec{y}$-direction
\begin{align}
\Delta \vec m \propto m_0\uvec{z} \times B_x\uvec{x} \propto m_0B_0^2\uvec{y}.\label{eq:dmy_osot}
\end{align} 
and changes the magnetization by about \SI{50}{\percent} as shown in Fig.~\ref{Ferro_fig1}\,(b). 
According to our theory in Eq.~\eqref{induced_OSOF}, the right and left circular polarization will induce an OSOF along the $+\uvec{x}$ and $-\uvec{x}$ directions respectively.
Therefore, the right circular pulse exerts a torque along $\uvec{y}$ (viz. $\uvec{z}\times\uvec{x}=\uvec{y}$) and similarly, the left circular pulse exerts a torque along $-\uvec{y}$. 
The change in magnetization is also antisymmetric in the helicity of the light pulse, similar to ZT effects in $\Delta m_y$.
Note that we have assumed that the material dependent parameter $\zeta$ is zero, which dictates that the antisymmetric behavior is not universal. 
In fact, the parameter $\zeta$ could be different for right and left circular pulse, meaning that the antisymmetry of the plots above would be broken \cite{Berritta2016}.

The effective contributions of the combined ZT and OSOT have been computed in Fig.~\ref{Ferro_fig1}\,(c). 
We note that the magnetization change in $m_y$ is opposite in helicity for ZT and OSOT (e.g., see fifth row plots in Figs.~\ref{Ferro_fig1}\,(a) and \ref{Ferro_fig1}\,(b)). 
Therefore, the final effective contribution becomes only about 5\,\% change in $\Delta {m}_y$. 
For the $z$-component of magnetization, the effective $\Delta {m}_z$ is negligibly small, even though individual changes are about 5\,\% due to ZT and OSOT.    

\begin{figure*}[t!]
\centering
\includegraphics[width = 1\textwidth]{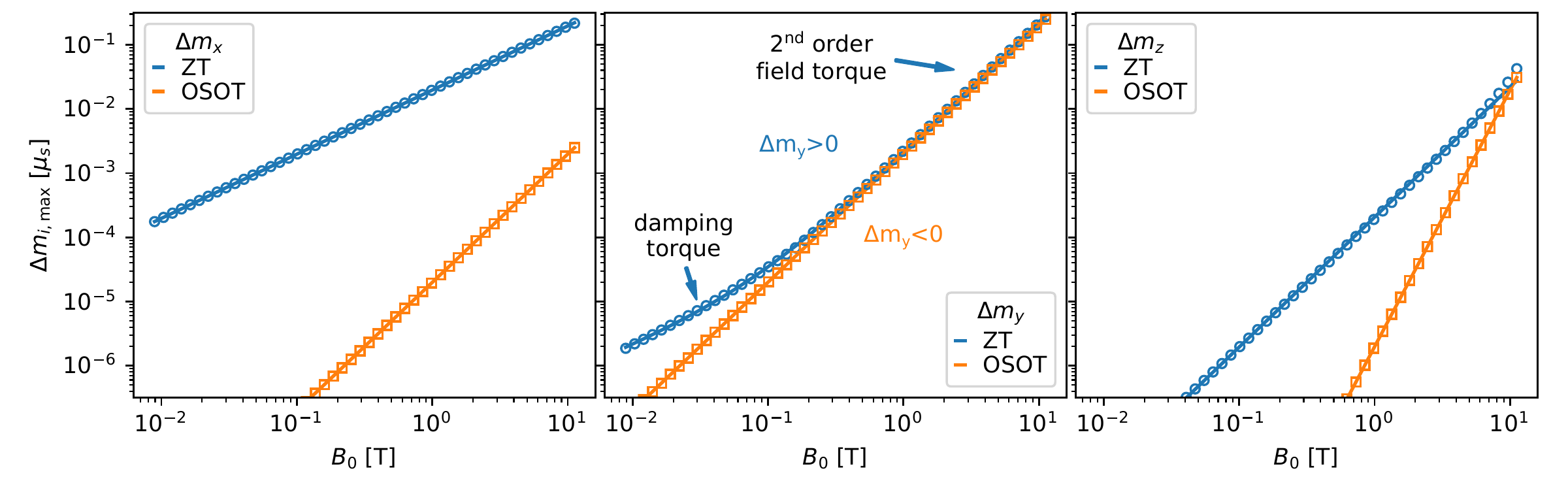}
\caption{(Color Online) Maximum of the magnetization change as a function of applied Zeeman field for the application of circular polarised THz pulses (absolute values). 
Lines are fits to the data according to a single (double) power law, see Tab.~\ref{tab:fits} for coefficients.} 
\label{Ferro_fig2}
\end{figure*}

To quantitatively understand the ZT and OSOT effects, we compute the field dependent contributions in terms of the maximum change of each magnetization components $\Delta m_i$ as a function of the applied field $B_0$ for left circular polarised light, shown in Fig.~\ref{Ferro_fig2}.
From a simple scaling argument we would expect the spin excitation $\Delta \vec m_\perp = \Delta m_x\uvec{x} + \Delta m_y\uvec{y}$ scaling with the magnitude of $\vec B_\text{EM}$ and $\vec B_\text{OSOF}$, respectively.
For the EM field, this is simply the amplitude $\vec B_\text{EM}\propto B_0$, whereas for the OSOF, that is $\vec B_\text{OSOF}\propto B_0^2$ according to Eq. (\ref{induced_OSOF}).

As already described above, Eq. \eqref{eq:dmx}, the spin excitation $\Delta m_x$ for the EM field is induced by the field-like torque from its $B_y$ component, thus, Fig.~\ref{Ferro_fig2} (a) shows a linear scaling with the EM field amplitude $B_0$.
The $\Delta m_y$ component in Fig.~\ref{Ferro_fig2} (b), however, is induced in a second order process, described in Eq. \eqref{eq:dmy} and thus scales quadratically with the EM field amplitude $B_0$, i.e., $\Delta m_y\propto B_0^2$.
A deviation of this scaling law is observed at lower field amplitudes of $B_0\lesssim\SI{100}{\milli\tesla}$, where the damping-like torque of the EM field with $\Delta m_y\propto \alpha B_0$ is taking over.
This effect can also be seen by comparing the torque amplitudes, which differ by a factor of $\Delta m_y/\Delta m_x\approx\alpha$ in this regime.
The $\Delta m_z$ component, i.e., the deviation of the equilibrium component then follows from conservation of magnetization amplitude: $\Delta m_z = m_0 - \sqrt{m_0-\Delta m_\perp^2}$. 
At low excitation the dominant contribution here is the $\Delta m_x$ component and one can simplify $\Delta m_z\approx-\Delta m_x^2/2m_0\propto -B_0^2$.

In contrast to the EM field, there is only one field-like torque acting along the $+\uvec{y}$ direction due to the OSOF in $\uvec{x}$ direction.
This torque can be seen in Fig.~\ref{Ferro_fig2} (b), and shows the quadratic scaling $\Delta m_y\propto -B_x\propto-B_0^2$ implied by Eq. \eqref{eq:dmy_osot}.
The $\Delta m_x$ excitation in Fig.~\ref{Ferro_fig2} (a) on the other hand is due to the damping-like OSOT and thus following the same power law, but a factor of $\alpha$ smaller compared to the field-like excitation in Fig.~\ref{Ferro_fig2} (b).
Finally, the $\Delta m_z$ excitation of the OSOT follows again from conservation of magnetization amplitude, leading to $\Delta m_z\approx-\Delta m_y^2/2m_0$ and implying $\Delta m_z\propto -B_0^4$ as shown in Fig.~\ref{Ferro_fig2} (c).

These scaling observations are once more summarized in Tab. \ref{tab:fits} where we display the scaling exponents obtained by fitting our simulation data. 
An interesting observation here is that although the characteristic field of the OSOT is on the order of \SI{100}{\tesla} in this frequency regime, due to the different torque symmetry compared to the ZT, the OSOT is by no means negligible.
For the $\Delta m_y$ excitation we find that the strength of ZT and OSOT are comparable, though opposite in sign, at fields on the order of \SI{e2}{\milli\tesla}---a value that is in much closer reach experimentally.
Altogehter, the OSOT effects can thus provide an equivalent torque compared to the Zeeman effect.
Therefore, the OSOT effects cannot be neglected when a circularly polarised light acts on a magnetic system even at the weak coupling limit $\zeta \rightarrow 0$. 

%To quantitatively understand the ZT and OSOT effects, we compute the field dependent contributions i.e., the maximum change of each magnetization components as a function of applied field  for left circular polarised light in Fig.~\ref{Ferro_fig2}. 
%As we have noted in Fig.~\ref{Ferro_fig1}, the magnitude of $\Delta {m}_y$ is equivalent for ZT and OSOT, however, they act in the opposite direction. 
%This is different for $\Delta{m}_z$, where the change is still equivalent in magnitude, but acting in the same direction. 
%To understand the plot for maximum of $\Delta{m}_x$ in Fig.~\ref{Ferro_fig2}, we note that the ZT consists of two torques.
% In fact, one torque directly points towards -$\uvec{x}$ direction, which explains the continuous decrease of the maximum of $\Delta{m}_x$ with increasing applied field for the ZT in the blue curve.
% In contrary, there is only one torque acting  along the $\uvec{y}$ direction due to OSOT, which helps the spins to rotate along the same direction.
% However, there is always an anisotropy field that acts along $\uvec{z}$. 
%The combination of these two torques results in small decrease of the maximum of $\Delta{m}_x$ with applied field.
%  As soon as the OSOT becomes larger (e.g., the applied field is large) than the anisotropy field, the competition is won by the OSOT.

Up to now we did not take into account the role of a finite anisotropy.
In order to investigate this, we performed the same calculation depicted in Fig.~\ref{Ferro_fig1} with the anisotropy $d_z = \SI{7.659}{\micro\electronvolt}$ for Fe included.
We found that the main effect of the anisotropy field is the precession of the induced magnetization $\Delta \vec m_\perp$ around the $z$-axis over time (see Appendix \ref{appendixB} for details).
On the other hand, no direct effect of the anisotropy can be observed on the ultrafast time scales, i.e., on the time scale of the pulse duration.

\begin{table}[!bt]
\caption{Fit parameters to a scaling function $f_N(B_0) = \sum_{n=1}^Na_{y,n} (B_0/\si\tesla)^{\beta_{i,n}}$ where $a_{i,n}$ is given in units of $\mu_s$.
For the ZF fit of $\Delta m_y$ a double power law $N=2$ has been used, whereas for the remaining fits a monomial $N=1$ was sufficient.}
\begin{ruledtabular}
\begin{tabular}{rrcc|cc}
&&\multicolumn{2}{c|}{ZF} &\multicolumn{2}{c}{OSOF}\\
& $n$ & $a_{i,n}$ & $\beta_{i,n}$ & $a_{i,n}$ & $\beta_{i,n}$\\\hline
$\Delta m_x$ & 1 & \num[scientific-notation = true]{  0.01975924} & \num[]{  1.00306052} & \num[scientific-notation = true]{  0.00009898} & \num[]{  1.99910498}\\
$\Delta m_y$ & 1 & \num[scientific-notation = true]{  0.00088026} & \num[]{  0.95948487} & \num[scientific-notation = true]{  0.00197803} & \num[]{  1.99858149}\\
             & 2 & \num[scientific-notation = true]{  0.00174323} & \num[]{  2.09038341} & -                                              & -                   \\
$\Delta m_z$ & 1 & \num[scientific-notation = true]{  0.00020343} & \num[]{  2.03214753} & \num[scientific-notation = true]{  0.00000196} & \num[]{  3.99720447}\\
\end{tabular}
\end{ruledtabular}
\label{tab:fits}
\end{table}

%{\it Temperature dependence of OSOT:}
\subsection{Finite temperature simulations}
For the finite temperature simulations, the exchange parameters of \citet{Pajda2001} are used, where the first two nearest neighbors are strongly ferromagnetically, however, the third nearest neighbor is weakly antiferromagnetically coupled.
For these simulations, we also use the uniaxial anisotropy of $d_z = \SI{7.659}{\micro\electronvolt}$ along the $z$-axis, in order to align the magnetization.
Calculating the temperature dependence of the OSOT for ferromagnetic Fe, we use a simulation grid of 48$^3$ spins and add a stochastic field to the effective field in Eq.~\eqref{LL-equation}, in order to treat the thermal fluctuations \cite{nowak2007handbook}. 
The calculated Curie temperature of this system is $T_{\rm C}=\SI{1368}\kelvin$ and thus slightly higher than the true value of Fe. 
However, since the Curie temperature is the only temperature scale in our simulations, we can basically treat it as a free scaling parameter.

\begin{figure}[hbt!]
\centering
\includegraphics[width=\columnwidth]{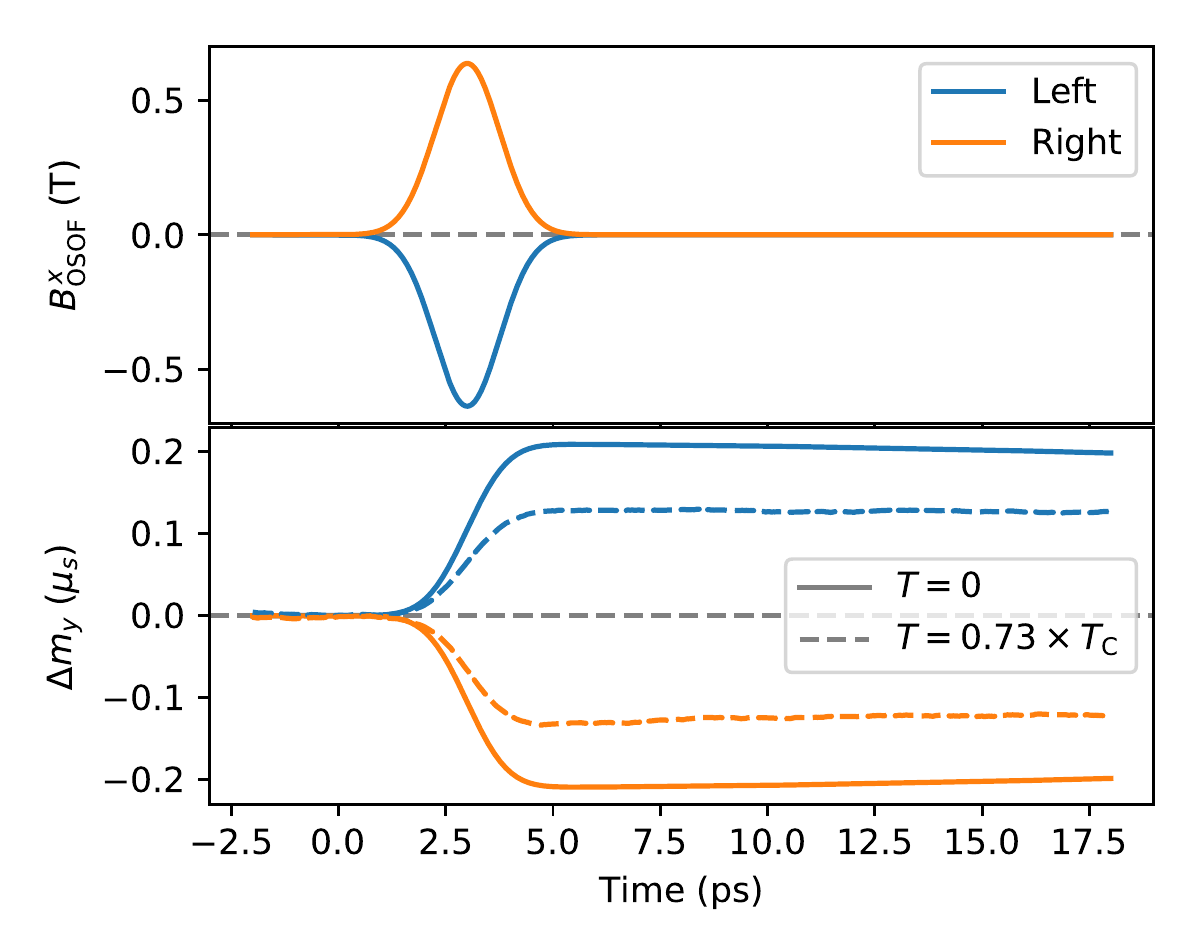}
\caption{(Color Online) OSOT effects at finite temperature. 
Top panel: the OSOT-induced field. Bottom panel: the corresponding spin dynamics for the  change of ${m}_y$ at $T=0$ and $T=\num{0.73}\times T_\text C$ (0\,K and 1000\,K). } 
\label{Ferro_fig3}
\end{figure}

\begin{figure}[hbt!]
\centering
\includegraphics[width=\columnwidth]{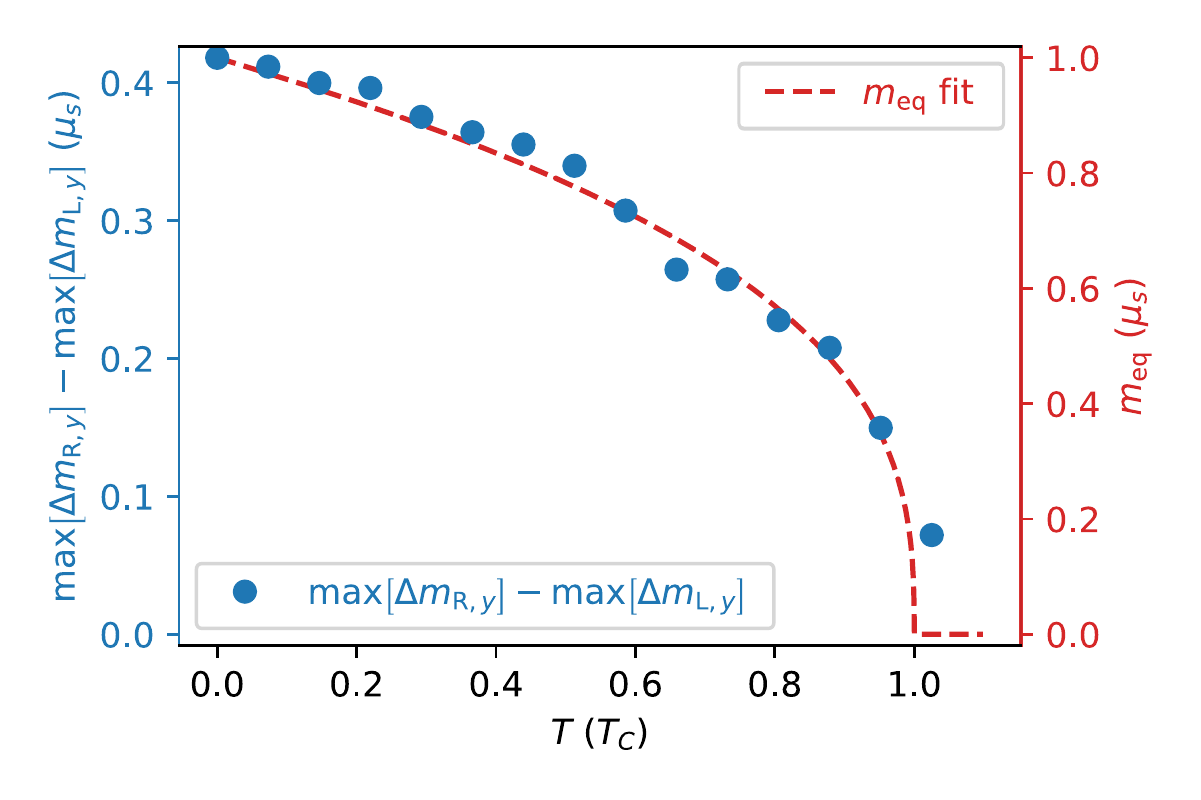}
\caption{(Color Online) Temperature dependence of OSOT, by the difference between the maximum changes of ${m}_y$ due to right and left polarised light pulses at maximum Zeeman field amplitude of \SI{10}{\tesla}.
For comparison, the right axis shows the equilibrium magnetization.} 
\label{Ferro_fig4}
\end{figure}

We compare in Fig.~\ref{Ferro_fig3} the OSOT effects at $T=0$ and $T=\num{0.73}\times T_\text C$ (0\,K and 1000\,K).
We find in Fig.~\ref{Ferro_fig4} that though the OSOT effect in $\Delta m_i$ appears to decrease with increasing temperature, this reduction is only related to the thermal reduction of magnetic order $m_\text{eq}(T)$ at finite temperature.
In other words the rotation angle of the normalized magnetization is not sensitive to the temperature.

% At 1000\,K, which is close to but below the Curie temperature of Fe, the OSOT still shows a significant effect. 

To illustrate this further, we have systematically calculated the temperature dependence of the OSOT by calculating the difference between maximum spin excitation for right and left circular pulses in Fig.~\ref{Ferro_fig4}. 
For each temperature, we performed ten simulations for each circular pulses, and took the average to determine $\operatorname{max}\left[\Delta {m}_{\text R,y}\right]  - \operatorname{max}\left[\Delta {m}_{\text L,y}\right]$ as a function of temperature, which ensures that the thermal fluctuations are minimised. 
Far away from $T_\text C$ the spin excitation amplitude is proportional to the equilibrium magnetization curve.
Only in the close vicinity of the critical temperature, we find an increase of the net spin excitation, relative to the equilibrium magnetisation.
This is simply due to the large amplitude of both ZF and OSOF which induce a transient magnetic order.

Therefore, OSOT effects should be observed for bcc Fe even at elevated temperatures i.e., at realistic conditions for ultra-intense spin excitations.
% Moreover, we observe that the OSOT effect does even not vanish at the Curie temperature. 

%{\it Conclusions}:
\section{Conclusions}
To conclude, we incorporated a new torque into the LLG equation that should appear in ultrafast spin dynamics, namely OSOT, and investigated this effect via computer simulations of an atomistic spin model, representative for bcc Fe.
%{\red How much of these statements are unique to this work and not \cite{Mondal2016}?}
The OSOT originates from the spin-orbit coupling of the electron spins to an external EM field. 
The strength of this OSOT, unlike the first-order ZT, depends on the \emph{intensity} of the ultrafast light pulse, as well as on the frequency.
In addition, the OSOT depends on the ellipticity of the pulse and it provides maximum torque for circularly polarised light pulses. 
%Moreover, the induced field is proportional to the coupling parameter ($\zeta$) of the light to the incident material. 
Throughout the simulations presented here, we considered the weak coupling limit $\zeta \rightarrow 0$. However, the coupling depends on the electronic configurations of the system and could potentially further increase the strength of the OSOT.
Although the OSOT is a higher order contribution in the external field, we found that the ZT and the OSOT provide quantitatively equivalent torques on the spins for circularly polarised laser pulses in the magnetization component perpendicular to the $\vec k$-vector and equilibrium magnetisation $\vec m_0$. 
%However, the effective ZT acts oppositely to the OSOT. 
The effect of the OSOT resembles an already known effect, namely the IFE and it can be considered as a relativistic contributions to the IFE. 
The temperature dependence study of OSOT shows that the OSOT effect is present at elevated temperatures, even up to the Curie temperature.
% The latter suggests that if sufficiently strong magnetic field amplitudes could be achieved, the OSOT effect could even be observed for magnetic systems above its critical temperature.

\section*{ Acknowledgments}
We thank  L\'aszl\'o Szunyogh for valuable discussions and acknowledge financial support from the Alexander von Humboldt-Stiftung, Zukunftskolleg at Universit\"at Konstanz via grant No. P82963319 and the Deutsche Forschungsgemeinschaft via NO 290/5-1.

 \appendix
 
 \section{Derivation of optical spin-orbit torque}
 \label{appendixA}
 
 Following Eq.~(\ref{osoc}), the induced field can be written as 
 \begin{align}
 \bm{B}_{\rm OSOT} & = -\frac{e^2\hbar(1+\zeta)}{4m^{2}c^{2}\mu_s} (\bm{E}_{\rm ext}\times\bm{A})\,.
 \label{B_osoc}
 \end{align} 
 We use the time dependent field as $\bm{E}_{\rm ext} = \mathcal{R}\left(\bm{E}_0 e^{i(\bm{k}\cdot\bm{r} - \omega t)}\right)$ and the amplitude as the elliptically polarised light $\bm{E}_0 = \frac{E_0}{\sqrt{2}} \left(\uvec{y} + \mathrm e^{i\eta}\uvec{z}\right)$, with $\eta$ as the ellipticity of the light. 
 Therefore, the electric field can be written as (when only the time-dependent part is taken) $\bm{E}_{\rm ext} = \frac{\partial \bm{A}}{\partial t} \Rightarrow \bm{A} = -\int \bm{E}_{\rm ext} \,dt$ that can be calculated as follows
\begin{align}
    \bm{A} %& = -\int \bm{E}_{\rm ext} \,dt\nonumber\\
     =-  \mathcal{R}\left[\bm{E}_0 \int e^{-i \omega t} dt\right] %= - \mathcal{R}\left[ \bm{E}_0 \frac{ e^{-i \omega t}}{-i \omega} \right] 
     = - \mathcal{R}\left[i  \frac{ \bm{E}_0e^{-i \omega t}}{ \omega} \right]\,.
\end{align}

Therefore, the induced field can be taken from Eq.~\eqref{B_osoc} as
\begin{align}
    \bm{B}_{\rm OSOT} & = \frac{e^2\hbar(1+\zeta)}{4m^{2}c^{2}\mu_s\omega}\mathcal{R}\left[i (\bm{E}_0\times \bm{E}_0^\star)\right]\nonumber\\
    %& = \frac{e^2\hbar(1+\zeta)}{4m^{2}c^{2}\mu_s\omega}\frac{E_0^2}{2}\mathcal{R}\left[i \left(\hat{\bm{e}}_y+ e^{i\eta}\hat{\bm{e}}_z\right)\times \left(\hat{\bm{e}}_y+ e^{-i\eta}\hat{\bm{e}}_z\right)\right]\nonumber\\
    %& = \frac{e^2\hbar(1+\zeta)}{4m^{2}c^{2}\mu_s\omega}\frac{E_0^2}{2}\mathcal{R}\left[i \left( e^{-i\eta}  - e^{i\eta} \right)\hat{\bm{e}}_x\right]\nonumber\\
    %& = \frac{e^2\hbar(1+\zeta)}{4m^{2}c^{2}\mu_s\omega}\frac{E_0^2}{2}\mathcal{R}\left[-i \left(  e^{i\eta} -e^{-i\eta}  \right)\hat{\bm{e}}_x\right]\nonumber\\
    %& = \frac{e^2\hbar(1+\zeta)}{4m^{2}c^{2}\mu_s\omega}\frac{E_0^2}{2}\mathcal{R}\left[-i \left(  2i\sin \eta \right)\hat{\bm{e}}_x\right]\nonumber\\
    & = \frac{e^2\hbar(1+\zeta)}{4m^{2}c^{2}\mu_s\omega} \,E_0^2\,\sin \eta\, \uvec{x}\nonumber\\
    & = \frac{e^2\hbar(1+\zeta)}{4m^{2}\mu_s\omega} \,B_0^2\,\sin \eta\, \uvec{x}
\end{align}
In the last step of the calculation, we used the relation  $E_0 = cB_0$. In our simulations, we apply time-dependent Zeeman fields along $y$ and $z$-directions and the corresponding induced optical spin-orbit field acts along $x$-direction.

\section{Effect of anisotropy}
\label{appendixB}

\begin{figure*}[hbt!]
\centering
\includegraphics[width =0.85 \textwidth]{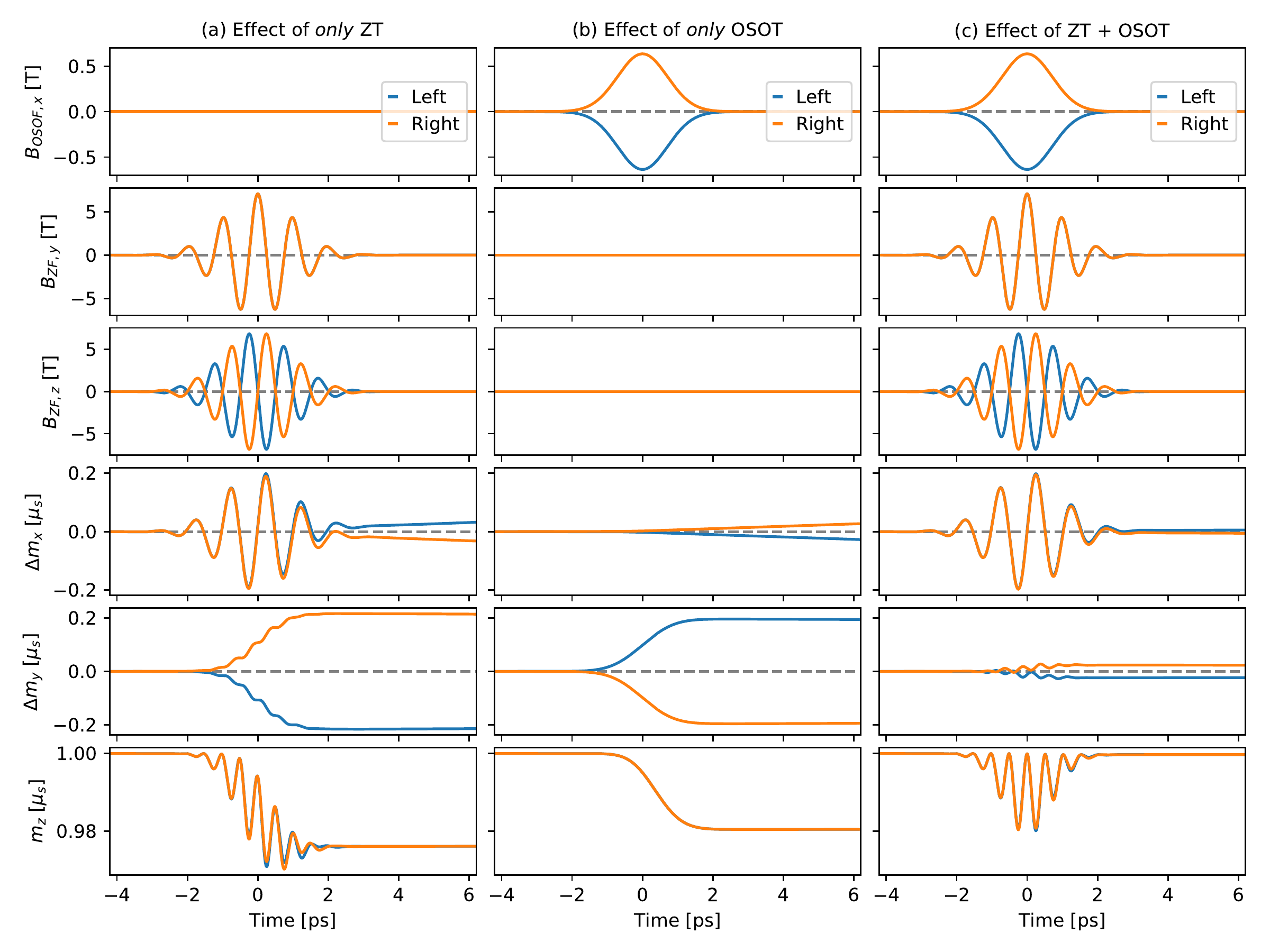}
\caption{(Color Online) The dynamical effects of ZT and OSOT have been calculated for Fe including the uniaxial anisotropy at an applied field of \SI{10}{\tesla}. 
The calculations have been performed accounting (a) {\it only} Zeeman effect, (b) {\it only} OSOT effect and (c) both the Zeeman and OSOT effects.
In all the plots, the first row represents the induced spin-orbit field which has only $x$-component, and the rows two and three show the applied ZF which has $y$ and $z$ components. 
The other three rows show the change in magnetization along $x$, $y$ and $z$ components respectively.
The orange and blue colors represent the action of right and left circular pulses.} 
\label{SM_fig1}
\end{figure*}

Here, we compute the influence of uniaxial magnetic anisotropy on the spin dynamics induced by the ZT and OSOT.
Fig.~\ref{SM_fig1} shows the magnetization dynamics taking into account the uniaxial anisotropy for bcc Fe.  
The main effect of anisotropy can be noticed by comparing with the zero-anisotropy simulations in Fig.~\ref{Ferro_fig1}.
A small increase in $\Delta m_x$ is observed in the case of only ZF or OSOF after the pulse has passed.
This is due to the slow precession of the induced $\Delta m_y$ component which starts to precess around the anisotropy field along $\uvec{z}$ axis.
In case of the superposition of ZF and OSOF the excitation of $\Delta m_y$ is mostly compensated and therefore no $\Delta m_x$ emerges either.
Additionally, we mention that the anisotropy energies do not affect the $\Delta m_y$ and $\Delta m_z$ on the ultrafast time scale and the net excitation remains the same irrespective of the anisotropy --- at least for the typically weak magnetic anisotropies of the 3d ferromagnets.

\bibliographystyle{apsrev4-1}
%\bibliography{References}

%\end{document}
\end{document}